\documentclass[12pt]{article}
\usepackage[pctex32]{graphics}
\begin{document}
\vspace{1cm}
\begin{center}
~\\
{\bf  \Large Thermodynamics on Fuzzy Spacetime}
\vspace{1cm}

                      Wung-Hong Huang\\
                       Department of Physics\\
                       National Cheng Kung University\\
                       Tainan, Taiwan\\

\end{center}
\vspace{1cm}
\begin{center}{\bf  \Large ABSTRACT } \end{center}
We investigate the thermodynamics of non-relativistic and relativistic ideal gases on the spacetime with noncommutative fuzzy geometry.   We first find that the heat capacities of the non-relativistic ideal boson and fermion on the fuzzy two-sphere have different values, contrast to that on the commutative geometry.  We calculate the ``statistical interparticle potential"  therein and interprete this property as a result that the non-commutativity of the fuzzy sphere has an inclination to enhance the statistical ``attraction (repulsion) interparticle potential" between boson (fermion).   We also see that at high temperature the heat capacity approaches to zero.  We next evaluate the heat capacities of the non-relativistic ideal boson and fermion on the product of  the 1+D (with D=2,3) Minkowski spacetime by a fuzzy two-sphere and see that the fermion capacity could be a decreasing function of temperature in high-temperature limit,  contrast to that always being an increasing function on the commutative geometry.  Also, the boson and fermion heat capacities both approach to that on the 1+D Minkowski spacetime in high-temperature limit.  We discuss these results and mention that the properties may be traced to the mechanism of  ``thermal reduction of the  fuzzy space".  We also investigate the same problems in the relativistic system with free Klein-Gordon field and Dirac field and find the similar properties. 
\vspace{1cm}
\\
\\
\begin{flushleft}
*E-mail:  whhwung@mail.ncku.edu.tw\\
\end{flushleft}


\newpage
\section{Introduction}
Physics on the noncommutative spacetime had been received a great deal of attention [1-8].  Historically, it is a hope that the deformed geometry in the small spacetime would be possible to cure the quantum-field divergences, especially in the gravity theory.  The renovation of the interesting in noncommutative field theories is that it have proved to arise naturally in the string/M theories [3,4]. In the noncommutative geometric approach [5] to the unification of all fundamental interactions, including gravity,  the space-time is the product of an ordinary Riemannian manifold M by a finite noncommutative space F.  The need for F is to avoid the fermion doubling problem [5]. An  advantage of this approach over the traditional grand unification approach is that the reduction to the Standard Model gauge group is  not due to plethora of Higgs fields, but is naturally obtained from the order one condition, which is one of the  axioms of noncommutative geometry [5]. 

  Motivated by the physical interesting of noncommutative spacetime we will in this paper study the thermodynamics of ideal gas on the spacetime which is the product of a 1+D Minkowski manifold by a noncommutative fuzzy geometry. Note that the noncommutative fuzzy sphere can appear naturally in the string/M theory [6,7].  It is known to correspond to the sphere D2-branes in string theory with background linear B-field [8]. Also, in the presence of constant RR three-form potential, the D0-branes can expand into a noncommutative fuzzy sphere configuration [9]. 

   In section II we first review the mathematical property of fuzzy sphere [10-11].   Then, to get some feelings about the thermal property on the fuzzy geometry we first evaluate the heat capacity of the non-relativistic ideal boson and fermion on the fuzzy two-sphere.  We see that they have different values, contrast to that on the commutative geometry [12].  In section III we calculate the ``statistical  interparticle potential"  [13] and see that  the noncommutativity of the fuzzy sphere has an inclination to enhance the statistical ``attraction (repulsion) interparticle potential" between boson (fermion).   This statistical  property may be used to explain why the ideal boson and fermion on the fuzzy two-sphere have different value of heat capacity. 

 In section IV we evaluate the heat capacities of the non-relativistic ideal boson and fermion on the product of  1 + D Minkowski spacetime by a fuzzy two-sphere, with D=2,3.  We find that the  heat capacities therein approach to those on the 1+D Minkowski spacetime in the high-temperature limit.  Also, the boson and fermion heat capacities become decreasing function of temperature in high-temperature limit, contrast to the property that fermion heat capacity is always an increasing function on the 1+D commutative geometry. Note that in the 1+2 commutative spacetime the boson and fermion have same heat capacity value which is an increasing function of temperature.  The calculations are performed in section 4 and summarized in table 1. We discuss these results and mention that the properties may be traced to the mechanism of  ``thermal reduction of the fuzzy space". 
\\
\\
{\hspace{1cm} {\it Table 1:  The high-temperature heat capacities of non-relativistic ideal  boson and fermion on the Minkowski spacetime + fuzzy two-sphere and that on the flat Minkowski spacetime.}
\\
\\
\scalebox{1}{\hspace{2cm}\includegraphics{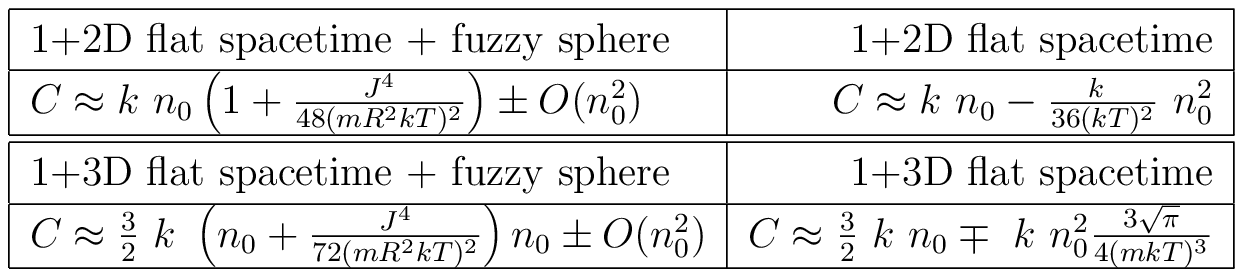}}
\\

  In section V we present a simple toy model to see such an interesting property.   In section VI we investigate the relativistic system and evaluate the heat capacity of the free Klein-Gordon and Dirac field on the product of  1 + 2 Minkowski spacetime by a fuzzy two-sphere.   We also find that the  heat capacity therein approaches to that on the 1+ 2 Minkowski spacetime in high-temperature limit.  Last section is devoted to a discussion.  Note that the properties of Casimir effect and effective potential on the noncommutative fuzzy space had been studied by us in [14].

\section{Thermodynamics of Non-relativistic Gas on Fuzzy Two-Sphere}
 The noncommutative fuzzy two-sphere geometry is described by a finite dimensional algebra generated by 3 matrices $X_i$ which satisfies  the commutator [10]
$$[ X_a, X_a] = i {R\over \sqrt{J(J+1)}}\epsilon_{abc} X_c, ~~~a,b,c = 1,2,3~~~~ and~~~2J\in N.\eqno{(2.1)} $$
in which $X_a$ are $(2J+1)  \times  (2J+1)$ matrices proportional to the $(2J+1)$-dimensional represent of the generators $\bf J$  of $SU(2)$ algebra.  It is known that in the limit $J \rightarrow \infty$ at fixed $R$ we get the ordinary sphere with radius $R$.  The non-relativistic free particle with mass $m$ on the fuzzy sphere has the spectrum [10]
$$ E_\ell= {\ell(\ell+1)\over 2mR^2},  ~~~~~\ell = 0\cdot\cdot\cdot J,\eqno{(2.2)} $$
with degeneracy $2\ell+1$, which will be used in the following calculations.
\subsection{Classical Statistics of  Non-relativistic Gas}
To proceed, let us first investigate the classical statistics in which the partition function is defined by
$$Z = \sum_{\ell=0}^{J}(2\ell+1) e^{-{\ell(\ell+1)\over 2mR^2kT}}.\eqno{(2.3)}$$
Using the Euler-Maclaurin summation formula [15]
$$\sum_{n=b}^{a}f(n) =\int_a^b dx f(x) + {1\over2} [f(a)+f(b)]+{1\over12} [f'(a)-f'(b)]+\cdot\cdot\cdot.\eqno{(2.4)}$$
the partition function in the high temperature has an approximation value
$$Z \approx 2mR^2kT\left[1- e^{-{J(J+1)\over 2mR^2kT}}\right] +(J+{2\over3}) e^{-{J(J+1)\over 2mR^2kT}} +\cdot\cdot\cdot.\eqno{(2.5)}$$
The associated mean energy $E$ and heat capacity $C$ are 
$$ E(T) = -{\partial \ell n Z\over\partial\beta}\approx {J(J+1)\over 4mR^2} - {J^2(J+1)^2\over 48m^2R^4}{1\over kT}.\eqno{(2.6a)}$$
$$ C(T) = {\partial E\over\partial T}\approx {J^2(J+1)^2\over 48m^2R^4}{1\over kT^2}.\hspace{2.7cm}\eqno{(2.6b)}$$
\\
The asymptotic value of energy $E(T\rightarrow \infty)$ is just the algebra mean value of (2.2) calculated by
$$  {\sum_{\ell=0}^{J}(2\ell+1) E_\ell\over \sum_{\ell=0}^{J}(2\ell+1) } ={J(J+1)\over 4mR^2}= E(T\rightarrow\infty).\eqno{(2.7)}$$
This is a finite value as the maximum energy of a particle has  a finite value $E_J$, in which $J$ is a finite value.  The finite asymptotic value $E(\infty)$  implies that  the heat capacity becomes zero asymptotically.  

  Note that we could not use the approximation result (2.6) to find the quantity in the ordinary sphere by taking the limit of $J \rightarrow \infty$.  This is because that the approximation adopted in there is suitable only under the condition ${J(J+1)\over 2mR^2kT} \ll 1$.  This property will be found in the following section.
\subsection{Non-relativistic Ideal Boson and Fermion}
The thermodynamics of ideal  boson and fermion  could be calculated from the following two relations 
$$ N = \sum_p{1\over z^{-1}e^{\beta\epsilon}\pm 1}; ~~~E = \sum_p{\epsilon\over z^{-1}e^{\beta\epsilon}\pm 1},\eqno{(2.8)}$$
where $z$ is  fugacity of the ideal gas, which is related to the chemical potential $\mu$ through the formula $z\equiv exp(\mu/kT)$ [12]. Using the spectrum (2.2) the numerical results of the energy for  ideal boson and fermion are plotted in figure 1.
\\
\\
\scalebox{1}{\hspace{5cm}\includegraphics{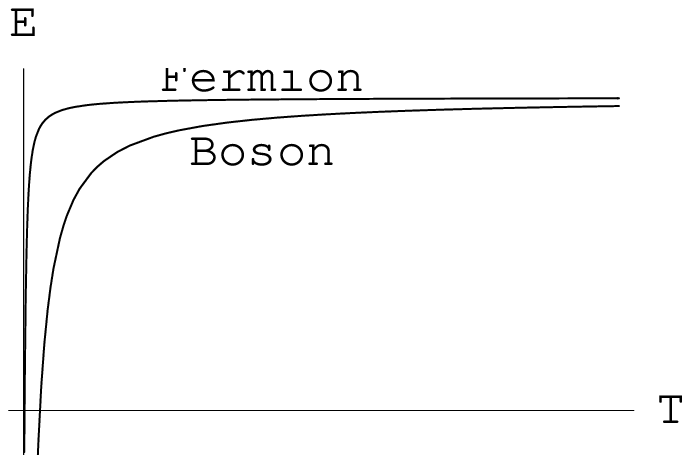}}
\\
\\
{\hspace{1cm} {\it Figure 1:  The energy of ideal  boson and fermion  on the noncommutative fuzzy sphere.}
\\
\\
Figure 1 shows that the heat capacities of the ideal boson and fermion on the fuzzy two-sphere have different values, contrast to that on the commutative geometry.  Also, as the gas have a finite maximum energy the associated heat capacity becomes zero asymptotically.  

\subsubsection{High-temperature Expansion}
  To see the above property we can perform the high-temperature expansion (i.e. ${J(J+1)\over 2mR^2}\ll kT$) to (2.8) with a help of  Euler-Maclaurin summation formula (2.4).  The results are

$$ N \approx z\left[2mkTR^2\left(1-e^{-{J(J+1)\over 2mR^2kT}}\right)+\left(J+{2\over3}\right)e^{-{J(J+1)\over 2mR^2kT}}+{1\over3}\right] \hspace{2.8cm}$$
$$~~~~\mp z^2\left[mkTR^2\left(1-e^{-{J(J+1)\over mR^2kT}}\right)+\left(J+{2\over3}\right)e^{-{J(J+1)\over mR^2kT}}+{1\over3}\right] + O(z^3).\hspace{1cm}\eqno{(2.9)}$$

$$ E \approx{1\over 2mR^2} \left[z\left((2mR^2kT)^2\left(1-e^{-{J(J+1)\over 2mR^2kT}}\right)-2mkTR^2J(J+1)e^{-{J(J+1)\over 2mR^2kT}}\right) \right.$$
$$\hspace{2.5cm}\left.\mp z^2\left((mR^2kT)^2\left(1-e^{-{J(J+1)\over mR^2kT}}\right)-mkTR^2J(J+1)e^{-{J(J+1)\over mR^2kT}}\right)\right] + O(z^3).\eqno{(2.10)}$$
In the case of $1\ll J(J+1)\ll 2mR^2kT$ we can use (2.9) to express the fugacity $z$ as a function of $N$.  After substituting this relation into (2.10) we find that the relation between the energy density  and  number density  becomes
$$ \varepsilon \approx  \left({J^2\over4mR^2} -{J^4\over24mR^2}{1\over 2mR^2kT} \right) n_0 \pm{J^2  \over6m}{1\over 2mR^2kT}~ n_0 ^2,\eqno{(2.11)}$$
which shows a different behavior between  the heat capacity of the ideal boson and fermion on the fuzzy two-sphere.   Note that, as the boson and fermion gas have a finite maximum energy the associated heat capacity becomes zero asymptotically.  Above equation is consistent with (2.6a) when $J\gg 1$.
\subsubsection{Low-temperature Expansion}
  In the low temperature the fugacity does not approach to zero and we need to adopt another approach.  For the case of fermion gas we can first use the Euler-Maclaurin summation formula (2.4) to express the particle number as
$${N} = \sum_{\ell=0}^J{2\ell +1\over z^{-1}e^{-{J(J+1)\over 2mR^2kT}}+ 1}
 \approx {2mR^2\over \beta} \left(- \ell n(1+ ze^{-{J(J+1)\over 2mR^2kT}}) +\ell n(1+z)\right)$$
$$\hspace{4cm}+{1\over2}{1\over 1+z^{-1}}+{1\over2}{2J+1\over1+z^{-1}e^{{J(J+1)\over 2mR^2kT}}}.\eqno{(2.12)}$$
In the low temperature the fugacity become infinite and above relation can be approximated as
$$N \approx  {2mR^2\over \beta} \left(- ze^{-{J(J+1)\over 2mR^2kT}} +\ell n z \right)+{1\over2}\left({1-z^{-1}}\right)+{1\over2}\left(2J+1\right) ze^{-{J(J+1)\over2mR^2kT}},\eqno{(2.13)}$$
which implies that 
$$ z \approx e^{N\over 2mR^2kT} - J e^{{-J(J+1)\over 2mR^2kT}}.\eqno{(2.14)}$$
Note that substituting above result into the last term in (2.13) we see that 
$$ze^{-{J(J+1)\over2mR^2kT}} \approx e^{N-J(J+1)\over 2mR^2kT} - J e^{{-2J(J+1)\over 2mR^2kT}} \rightarrow 0,\eqno{(2.15)}$$
at low temperature, as the total particle number $N$ shall be less then the total acceptable state $N_{max}\equiv\sum_{\ell=0}^J 2\ell+1$.  Thus the low temperature expansion in (2.13) is a consistent method.

Next, we also use the Euler-Maclaurin summation formula (2.4) to express the particle energy  as
$${E} = \sum_{\ell=0}^J{2\ell +1\over z^{-1}e^{-{J(J+1)\over 2mR^2kT}}+1} {\ell(\ell+1)\over 2mR^2}\approx {2mR^2(kT)^2} \left({- \sf Li_2}(-z^{-1}e^{{J(J+1)\over 2mR^2kT}}) + {\sf Li_2}(- z^{-1}) \right)$$
$$\hspace{4cm}- kT \ell n(1+ z^{-1}e^{{J(J+1)\over 2mR^2kT}}) +{1\over 2mR^2}\left({J^4\over2}+{J^3\over 1+z^{-1}e^{{J(J+1)\over 2mR^2kT}}}\right),\eqno{(2.16)}$$
in which ${\sf Li_2}(\mp y)$ is the polylogarithm function which has a series expansion formula [15]
$${\sf Li_2}(y) = \sum_{k=1}^\infty  {y^2\over k^2}.\eqno{(2.17)}$$
In the low temperature, as the fugacity $z \rightarrow \infty$ the term ${\sf Li_2}(- z^{-1})$ in (2.17) could be expressed a series expansion.   However, as $ z^{-1}e^{{J(J+1)\over 2mR^2kT}}\rightarrow \infty$ (as explained in (2.15)) we have to use the ``inversion formula" [15]
$${\sf Li_2}(y)+ {\sf Li_2}(y^{-1}) = -{\pi^2\over 6}- {1\over2}\left(\ell n \left(-y\right)\right)^2,\eqno{(2.18)}$$
to make a series expansion about another polylogarithm function in (2.16).  After a lengthy algebra evaluations we find that the energy density becomes
$$ {E\over 4\pi R^2} ={1\over 2m} \left({n_0^2\over2} +{\pi^2\over 6} (kT)^2\right) +O(e^{-J^2/R^2kT}),\eqno{(2.19)}$$
in which ${n_0}$ is the particle number density.   Above result is just the relation in the commutative system with small correction $O(e^{-J^2/R^2kT})$.

Note that at zero temperature the particle will filled from the state $\ell =0$ to $\ell= J$.   In the case of $J \gg 1$ the total particle number and associated energy are
$$ N =\sum_{\ell=0}^{J} \approx J^2,~~~~~E =\sum_{\ell=0}^{ J}{\ell(\ell+1)\over 2m} \approx {1\over 2m} {J^4\over2},~~~\Rightarrow E = {1\over 2m} {N^2\over2}\eqno{(2.20)}$$
which precisely give the leading term in (2.19).   
\\

   Let us present the physical interpretation about these thermal properties to conclude this section. 

  1. At high energy, as the system has a maximum value in the spectrum $ E_J= {J(J+1)\over 2mR^2}$ the system therefore has a finite  limiting energy as shown in (2.6a) and (2.11).  This implies that the heat capacity becomes zero asymptotically.  

 2.  At low temperature, as the particles are at low energy level they does not feel the constraint property of  $\ell\le J$, the system will behave as that on the commutative space, as shown in (2.19).

3. It is well known that, in comparison with the normal statistical behavior, bosons exhibit a larger tendency of bunching together, i.e., a positive statistical correlation. In contrast, fermions exhibit a negative statistical correlation. Uhlenbeck [13] interpreted this property by the ``statistical interparticle potential".   In our model, particle on the fuzzy sphere will be constrained  between the state with quantum number $\ell=0$ and $\ell=J$.   Therefore the fermion will feel more statistic repulsive effect and the boson will feel more statistic attractive effect, as shown in the next section.   The extra  statistical effect, which is induced by the fuzzy property, will render the  the heat capacities of the ideal boson and fermion on the fuzzy two-sphere to have different values, contrast to that on the commutative geometry,  as shown in (2.11).

\section{Statistical Interparticle Potential}
We now following the Uhlenbeck [12,13] to evaluate the ``statistical interparticle potential"  for the non-relativistic ideal boson and fermion on the fuzzy two-sphere.

  Define the one particle  matrix element of the Boltzmann factor by 
$$F_{ij} = <X_i|e^{-\beta H}|X_j>,\eqno{(3.1)}$$ 
then the matrix element of the Boltzmann factor for a system of two identical particles can be written as 
$$ <X_1,X_2|e^{-\beta H}|X_1,X_2>= F_{11}F_{22}\pm F_{12}F_{21},\eqno{(3.2)}$$ 
where the plus (minus) sign is adopted for the boson (fermion) system. For a translation symmetry system $F_{11}=F_{22}$ and $F_{12}=F_{21}$
and density matrix element becomes [12]
$$<X_1,X_2|\tilde \rho|X_1,X_2>= {1\over V^2}\left[ 1\pm {F_{12}^2\over F_{11}^2}\right], \eqno{(3.3)}$$ 
in which we define the density matrix by $\tilde \rho \equiv {e^{-\beta H}\over \sf {Tr}~e^{-\beta H}}$ [12] and $V$ is the system volume.  The ``statistical interparticle potential" $U$ is defined to be such that the Boltzmann factor exp$(-\beta v)$ is precisely equal to the correlation factor (bracket term) in the above equation, i.e., 
$$U = -kT\ell n \left[1\pm {F_{12}^2\over F_{11}^2}\right]. \eqno{(3.4)}$$

   Now, as the particle wave function on the fuzzy sphere is spherical harmonics $Y_\ell^m(\theta,\phi)$ [10] with $\ell \le J$,  we find that 
$$  F_{11} = Tr <X_1|e^{-\beta H}|X_1> = \sum_p <X_1|p><p|e^{-\beta H}|p><p|X_1>\hspace{4.5cm}$$
$$ = \sum_\ell\sum_{m_1}\sum_{m_2}\int d\phi d\theta~\sin\theta~ Y_\ell^{*m_1}(\theta,\phi)Y_\ell^{m_2}(\theta,\phi)~e^{-{\ell(\ell+1)\over 2mR^2kT}} = \sum_{\ell=0}^J (2\ell+1)\ell ~e^{-{\ell(\ell+1)\over 2mR^2kT}}.\eqno{(3.5)}$$
$$  F_{12} = <X_1|e^{-\beta H}|X_2> = \sum_p <X_1|p><p|e^{-\beta H}|p><p|X_2>\hspace{5.5cm}$$
$$ = \sum_\ell\sum_{m_1}\sum_{m_2}~ Y_\ell^{*m_1}(\theta_1,\phi_1)Y_\ell^{m_2}(\theta_2,\phi_2)~e^{-{\ell(\ell+1)\over 2mR^2kT}}.\eqno{(3.6)}$$

Substituting (3.5) and (3.6) into (3.4) the  ``statistical interparticle potential"  
is plotted  in figure 2.  The dashed line represents that with $J=5$ while solid line is that with $J=12$. 
\\
\\
\scalebox{1}{\hspace{5cm}\includegraphics{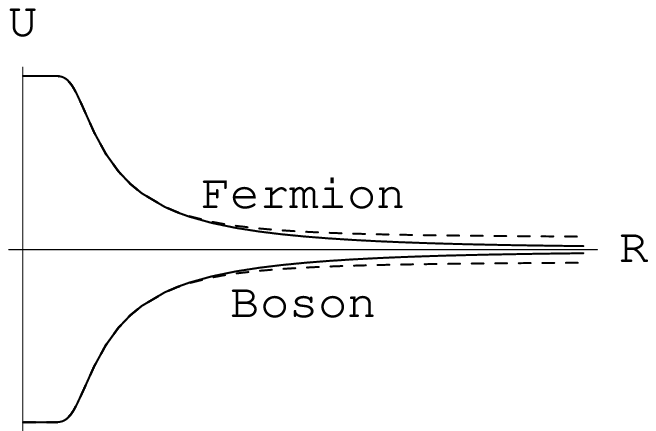}}
\\
\\
{\hspace{1cm} {\it Figure 2:  ``Statistical interparticle potential"  U(R) on the fuzzy geometry.  Dashed lines represents that with $J=5$ while solid line is that with $J=12$.}
\\
\\
Above results show that the fuzzy of sphere with finite value of $J$ will enhance the negative statistical correlation between fermion and enhances the positive statistical correlation between bosons.  Thus the thermal property of the boson and  fermion gas on fuzzy sphere will have different heat capacity, contrast to that on the commutative geometry. 

\section{Thermodynamics of ~Non-relativistic Gas on 1+D Minkowski Spacetime with Extra Fuzzy Sphere}
We now investigate the  thermodynamics of boson and fermion on 1+D Minkowski spacetime with extra fuzzy sphere.  We will see that at high temperature the thermodynamics of ideal gas will behave as that on the 1+D Minkowski spacetime without the extra fuzzy sphere.   This ``mechanism of thermal reduction of fuzzy geometry" will qualitatively modify the heat capacity of the gas.   
\subsection{Thermodynamics on 1 + 2  Minkowski Spacetime with Extra Fuzzy Sphere}
In the high-temperature limit $(i.e~ {J(J+1)\over 2mR^2}\ll kT$) the thermodynamics of ideal  boson and fermion on the 1+2  Minkowski spacetime with extra fuzzy sphere could be studied from the following analysis.
$$ N ={2\pi S\over h^2}\sum_{\ell=0}^J (2\ell +1)\int dp~ p{1\over z^{-1}e^{{\ell(\ell+1)\over 2mR^2kT}+{p^2\over 2mkT}}\pm 1} \approx z\left[{2\pi S\over h^2} \int dp~p~e^{-p^2\over 2mkT}\right]\sum_{\ell=0}^J (2\ell +1) e^{-\ell(\ell+1)\over 2mR^2kT}\hspace{2cm} $$
$$\mp z^2\left[{2\pi S\over h^2} \int dp~p~ e^{-p^2\over mkT}\right]\sum_{\ell=0}^J (2\ell +1) e^{-\ell(\ell+1)\over mkT}\approx {2\pi S\over h^2}2mkT\left[z\cdot W1 \mp z^2\cdot W2 \right].\eqno{(4.1)}$$
$$ E ={2\pi S\over h^2}\sum_{\ell=0}^J (2\ell +1)  \int dp~ p~{{p^2\over2m}+{\ell(\ell+1)\over2mR^2}\over z^{-1}e^{{\ell(\ell+1)\over 2mR^2kT}+{p^2\over 2mkT}}\pm 1}\approx {2\pi S\over h^2}\left[2mkT(z\cdot W1\mp z^2\cdot W2) \right.\hspace{2.5cm}$$
$$\hspace{7cm}\left.+(2mkT)^2 (z\cdot W3 \mp z^2\cdot  W4)\right],\eqno{(4.2)}$$
in which $S$ is the area of Minkowski space and we have defined
$$W1\equiv 2mkTR^2 \left(1-e^{-J(J+1)\over 2mR^2kT}\right)+\left(J+2/3\right)e^{-J(J+1)\over 2mR^2kT}+1/3.\hspace{1cm}\eqno{(4.3)}$$
$$W2 \equiv mkTR^2 \left(1-e^{-J(J+1)\over mR^2kT}\right)+\left(J+2/3\right)e^{-J(J+1)\over mR^2kT}+1/3.\hspace{1cm}\eqno{(4.4)}$$
$$W3\equiv (2mkTR^2 )^2\left(1-e^{-J(J+1)\over 2mR^2kT}\right)- 2mkTR^2J\left(J+1\right) e^{-J(J+1)\over 2mR^2kT}.\eqno{(4.5)}$$
$$W4\equiv (mkTR^2 )^2\left(1-e^{-J(J+1)\over mR^2kT}\right)- mkTR^2J\left(J+1\right)e^{-J(J+1)\over mR^2kT}.\hspace{0.4cm}\eqno{(4.6)}$$
Using (4.1) we can express the fugacity $z$ as a function of $N$.  After substituting this relation into (4.2) we can find the energy.  The associated heat capacity is
$$ C\approx  k~ n_0\left( 1 + {J^4\over 48(mR^2kT)^2}\right) \pm O(n_0^2),~~~(gas~on~1+2~flat~space~+fuzzy~sphere)  ,\eqno{(4.7)}$$
in which ${n_0}$ is the particle number density.  Above result shows that the heat capacity  is a decreasing function of temperature.  Note that in the 1+2 commutative spacetime the heat capacity of  boson and fermion has a same value
$$ C \approx  k~ n_0  - {k~ \over 36 (kT)^2}~n_0^2,~~~(gas~on~1+2~flat~space),\eqno{(4.8)}$$
which is an increasing function of temperature.   

Above result may be interpreted  as following.    At low temperature, as the particles are at low energy level they does not feel the constraint  property of $\ell\le J$ the system will behave as that on the 1+2+2 commutative space.  However, at high temperature, as the system has a maximum value in the spectrum $ \ell \le J$ the heat capacity of the system therefore behave as that on the 1+2 commutative space asymptotically.   This ``mechanism of thermal reduction of fuzzy geometry" will render the heat capacity of  boson and fermion to be  a decreasing function at high temperature.     

\subsection{Thermodynamics on 1+3  Minkowski  Spacetime with Extra Fuzzy Sphere}
The high-temperature thermodynamics of ideal  boson and fermion on the 1+3  Minkowski  spacetime with extra fuzzy sphere could be studied from the following analysis.
$$ N ={4\pi V\over h^3}\sum_{\ell=0}^J  (2\ell +1) \int dp~ p^2{1\over z^{-1}e^{{\ell(\ell+1)\over 2mR^2kT}+{p^2\over 2mkT}}\pm 1}\approx z\left[{4\pi V\over h^3} \int dp~ p^2 e^{-p^2\over 2mkT}\right]\sum_{\ell=0}^J (2\ell +1)  e^{-\ell(\ell+1)\over 2mR^2kT}\hspace{4cm}$$
$$ \mp z^2\left[{4\pi V\over h^3} \int dp~ p^2 e^{-p^2\over mkT}\right]\sum_{\ell=0}^J  (2\ell +1) e^{-\ell(\ell+1)\over mR^2kT}= {4\pi V\over h^3}{\sqrt\pi(2mkT)^{3/2}\over4}\left[z\cdot W1 \mp z^2\cdot  W2\right] .\eqno{(4.9)}$$
$$ E ={4\pi V\over h^3}\sum_{\ell=0}^J  (2\ell +1) \int dp~ p^2{{p^2\over2m}+{\ell(\ell+1)\over2mR^2}\over z^{-1}e^{{\ell(\ell+1)\over 2mR^2kT}+{p^2\over 2mkT}}\pm 1}\hspace{9cm}$$
$$\approx {4\pi V\over h^3}\left[{3\sqrt\pi(2mkT)^{5/2}\over8}(z\cdot W1\mp z^2\cdot W2) +{\sqrt\pi(2mkT)^{3/2}\over4} (z\cdot W3 \mp z^2\cdot  W4)\right],\eqno{(4.10)}$$
in which $V$ is the volume of Minkowski space.  W1, W2, W3 and W4 are defined in (4.3)-(4.6).  Now, using (4.9) we can express the fugacity $z$ as a function of $N$.  After substituting this relation into (4.10) we can find the energy.  The associated heat capacity is
$$ C\approx {3\over2}~k~\left(n_0 + {J^4\over 72(mR^2kT)^2}\right)n_0 \pm O(n_0^2),~~~(gas~on~1+3~flat~space~+fuzzy~sphere),\eqno{(4.11)}$$
in which ${n_0}$ is the particle number density.  Above result shows that the heat capacity is a decreasing function of temperature.   Note that in the 1+3 commutative spacetime the heat capacity of  boson and fermion has a same value
$$ C\approx {3\over2}~k~n_0 \mp ~k~n_0^2 {3\sqrt\pi\over 4(mkT)^3},~~~(gas~on~1+3~flat~space),\eqno{(4.12)}$$
which shows that  the heat capacity of boson is a decreasing function of temperature while that of fermion is an increasing function [12].  

  The physical interpretation is the same as that in previous subsection.  Thus the ``mechanism of  thermal reduction of the extra fuzzy space" could render a  particle to behave as that on lower commutative space, and it will have less heat capacity. The heat capacity, therefore will be a decreasing function at high temperature.   Note that in the 1+3  commutative space the boson has Bone-Einstein condensation, thus it has a  very large heat capacity  at transition temperature $T_c$.  Therefore, increasing the temperature beyond the  $T_c$ it will be a decreasing function.  

  More precisely, at low temperature the system does not feel the finite value property of $J$ and the particle will have the thermal property like as that on the commutative 1+3+2 commutative spacetime. In this case the heat capacity is an increasing function of temperature.  However, at high temperature the quantum level of  extra fuzzy space is all occupied and the particle will behave as that on lower space. Thus it will have less heat capacity.  This ``mechanism of  thermal reduction of the extra fuzzy space" could lead the heat capacity  to be a decreasing function of temperature and the heat capacity has  a ``peak value" near the temperature of `` thermal reduction".   The interesting property could be seen in the following toy model.
\section{A Toy Model}
 For example, let us consider a simplest toy model of  classical particle which  has spectrum 
$$E=a \cdot n + b\cdot \ell,  ~~~~~0\le n \le \infty,~~~~~0\le \ell \le J.\eqno{(5.1)}$$
The mode with quantum number $0\le n \le \infty$ is used to describe a simple harmonic oscillator and the mode with finite quantum number $0\le \ell \le J$ is used to simulate that on the fuzzy geometry.   In this model the partition function  and the heat capacity could be evaluated exactly.  We plot the heat capacity in figure 3.
\\
\\
\scalebox{1}{\hspace{5cm}\includegraphics{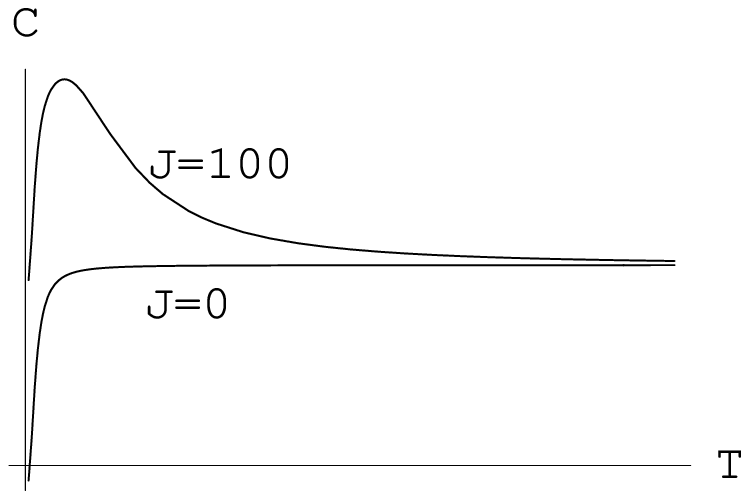}}
\\
\\
{\it Figure 3:  The heat capacity of the model with spectrum (5.1) when a=b=1.}
\\
\\
We have explicity seen that if $J \ne 0$ then there is a peak in the system capacity.
The associated heat capacity at low temperature is
$$ C \approx  {1\over kT^2}\left[a^2 e^{-a/kT}+b^2 e^{-b/kT}\right],~~~~low~ temperature,\eqno{(5.2)}$$
in which the quantum modes $n$ and $\ell$ contribute the similar behavior in heat capacity, which is an increasing function at low temperature.   On the other hand, at high temperature the associated heat capacity becomes
$$ C \approx  k+ {1\over 12kT^2}\left[b^2 J(J+2)-a^2\right],~~~~high~temperature.\eqno{(5.3)}$$
We see that the quantum mode $n$ contributes heat capacity value $ k-{a^2\over 12kT^2}$,  which is an increasing function,  and quantum mode $\ell$ contributes heat capacity value ${1\over 12kT^2}\left[b^2 J(J+2)\right]$, which is a decreasing function at high temperature.   Therefore, in the case of large $J$ with relation $b^2 J(J+2)>a^2$ the model will show a ``peak value" in  heat capacity, as shown in figure 3.
\section{Thermodynamics of  Relativistic  Gas  on Fuzzy Spacetime}
  We now turn to the problems with relativistic  gas.   The Klein-Gordon field equation on the product of  Minkowski  spacetime (with  coordinate $\vec x$) by a fuzzy two-sphere (with coordinate $J_i$) is
$$\left[\partial_t^2-\vec\nabla_{\vec x}^2 -J_1^2-J_2^2-J_3^2\right]\Phi + m^2\Phi =0.\eqno{(6.1)}$$
Expanding the scalar field as a product of plane wave $e^{-ip\cdot x}$ and spherical harmonic function $Y_m^{\ell}(\theta, \phi)$ then we see that 
the spectrum of  scalar field  is [10]
$$ E_\ell^2= \vec p^{~2} + {\ell(\ell+1)\over R^2} + m^2,  ~~~~~\ell = 0\cdot\cdot\cdot J,\eqno{(6.2)} $$
with degeneracy $2\ell+1$.  The Dirac field has a similar relation.  Note that the finite value of quantum number $\ell$ characterizes the fuzzy property of the fuzzy sphere. We will in following analyze the system will massless field for simplicity.
\subsection{Thermodynamics of  Relativistic Gas  on Fuzzy Two-Sphere}\subsubsection{Classical Statistics of  Relativistic Gas}
Let us first investigate the classical statistics for the relativistic gas  on fuzzy two-sphere.  In this case the partition function is defined by
$$Z = \sum_{\ell=0}^{J}(2\ell+1) e^{-{\sqrt{\ell(\ell+1)/R^2}\over kT}}.\eqno{(6.3)}$$
Using the Euler-Maclaurin summation formula in (2.4) the partition function in the high temperature (i.e. $ \sqrt {J(J+1)/ R^2} \ll kT$) has an approximation value
$$Z \approx 2R^2(kT)^2\left[1- \left(1+{\sqrt J\over RkT}\right)e^{-{\sqrt{\ell(\ell+1)/R^2}\over kT}}\right] +{1\over2}\left[1+\left(1+2J\right) e^{-{\sqrt{\ell(\ell+1)/R^2}\over kT}}\right]+\cdot\cdot\cdot.\eqno{(6.4)}$$
The associated mean energy $E$ and heat capacity $C$ are 
$$ E(T) = -{\partial \ell n Z\over\partial\beta}\approx {J\over 2R} - {J^2\over 4R^2}{1\over kT}.\eqno{(6.5a)}$$
$$ C(T) = {\partial E\over\partial T}\approx {J^2\over 4R^2}{1\over kT^2}.\hspace{1.7cm}\eqno{(6.5b)}$$
\\
The asymptotic value of energy $E(T\rightarrow \infty)$ is just the algebra mean value of (6.2) calculated by
$$  {\sum_{\ell=0}^{J}(2\ell+1) E_\ell\over \sum_{\ell=0}^{J}(2\ell+1) } ={J\over 2R}= E(T\rightarrow\infty).\eqno{(6.6)}$$
This is a finite value as the maximum energy of a particle has  a finite value $E_J$, in which $J$ is a finite value.  The finite asymptotic value $E(\infty)$  implies that  the heat capacity becomes zero asymptotically, as that in the non-relativistic system analyzed in section II. 
\subsubsection{Relativistic Ideal Boson and Fermion} 
The thermodynamics of  relativistic ideal boson and fermion field could be studied from the standard analysis [12] and we can perform the calculation from the (2.8).  Using the spectrum (6.2) and with a help of  Euler-Maclaurin summation formula (2.4) the results in high temperature are
$$ N \approx z\left[3J-\left({2J^{3/2}\over3RkT}+{\sqrt{J(J+1)}(1+2J)\over 2RkT}\right) \right]\mp z^2\left[3J+\left({4J^{3/2}\over3RkT}+{\sqrt{J(J+1)}(1+2J)\over RkT}\right) \right]  ,\eqno{(6.7)}$$
$$ E \approx z\left[\left({2J^{3/2}\over3R}+{\sqrt{J(J+1)}(1+2J)\over 2R}\right) +\left({J^2\over2R^2kT}-{{J(J+1)}(1+2J)\over 2R^2kT}\right) \right]\hspace{3cm}$$
$$\mp z\left[\left({2J^{3/2}\over3R}+{\sqrt{J(J+1)}(1+2J)\over 2R}\right) +\left({J^2\over R^2kT}-{{J(J+1)}(1+2J)\over R^2kT}\right) \right].\eqno{(6.8)}$$
We can use (6.7) to express the fugacity $z$ as a function of $N$.  After substituting this relation into (6.8) we find that the relation between the energy density $\varepsilon$ and  number density  $n_0$ becomes
$$ \varepsilon \approx  ~k~\left({J\over 2R}- {J^2\over 8R^2 kT}\right) n_0 \pm {J\over 16R^2 kT}~k~n_0^2,\eqno{(6.9)}$$
which shows a different behavior between  the heat capacity of the ideal boson and fermion.   Note that, as the boson and fermion gases have a finite maximum energy the associated heat capacity becomes zero asymptotically as that in the non-relativistic system which are analyzed in section II.  Above equation is consistent with (6.6).

\subsection{Thermodynamics of  Relativistic  Gas on 1+2  Minkowski  Spacetime with Extra Fuzzy Sphere}
To proceed, let us first remark that in order to have the analytic result we will investigate the system of massless free field on the 1+2 Minkowski spacetime with Extra Fuzzy Sphere. Note that, as the property we attempt to see is shown at high temperature, in which the quantum field will become asymptotic free and mass of the field is irrelevant.  Also, the property we find will be  shown in 1+3  Minkowski spacetime, after numerical analysis.

 Then, the thermodynamics of quantum Klein-Gordon field and Dirac field on the Kaluza-Klein spacetime of  ``1+2 + fuzzy sphere" could be studied from the standard analysis [12].  The total particle number $N$ and energy $E$  
could therefore be evaluated from the  two relations in (2.8).  Using the spectrum (6.2)  we find the following expressions in the high-temperature approximation (i.e. $ \sqrt {J(J+1)\over R^2} \ll kT$) 
$$ N \equiv{2\pi S\over h^2}\sum_{\ell=0}^J (2\ell +1)\int dp~ p{1\over z^{-1}e^{\beta \sqrt{p^2+{\ell(\ell+1)\over R^2}}}\pm 1} \hspace{7.2cm}$$
$$\approx {2\pi S\over h^2}\left[ z~\sum_{\ell=0}^J (2\ell +1)\int dp~p~{e^{-\beta \sqrt{p^2+{\ell(\ell+1)\over R^2}}}} \mp z^2~\sum_{\ell=0}^J (2\ell +1)\int dp~p~{e^{-2\beta \sqrt{p^2+{\ell(\ell+1)\over R^2}}}}\right]\hspace{0cm} $$
$$\approx {2\pi S\over h^2}\left[ z~\sum_{\ell=0}^J (2\ell +1)~{e^{-\beta \sqrt{{\ell(\ell+1)\over R^2}}}\over \beta^2}\left(1+ \beta \sqrt{{\ell(\ell+1)\over R^2}} \right)\right.\hspace{5.4cm}$$
$$\hspace{6cm} \left.\mp z^2~\sum_{\ell=0}^J (2\ell +1)~{e^{-2\beta \sqrt{{\ell(\ell+1)\over R^2}}}\over 4\beta^2}\left(1+ 2\beta \sqrt{{\ell(\ell+1)\over R^2}} \right)\right].\eqno{(6.10)}$$

$$ E \equiv{2\pi S\over h^2}\sum_{\ell=0}^J (2\ell +1)\int dp~ p{\sqrt{p^2+{\ell(\ell+1)\over R^2}}\over z^{-1}e^{\beta \sqrt{p^2+{\ell(\ell+1)\over R^2}}}\pm 1} \hspace{7.2cm}$$
$$\approx {2\pi S\over h^2}\left[ z~\sum_{\ell=0}^J (2\ell +1)\int dp~p~\sqrt{p^2+{\ell(\ell+1)\over R^2}}{e^{-\beta \sqrt{p^2+{\ell(\ell+1)\over R^2}}}}\right. \hspace{4.4cm}$$
$$\left. \mp z^2~\sum_{\ell=0}^J (2\ell +1)\int dp~p~\sqrt{p^2+{\ell(\ell+1)\over R^2}}{e^{-2\beta \sqrt{p^2+{\ell(\ell+1)\over R^2}}}}\right]\hspace{0cm} $$
$$\approx {2\pi S\over h^2}\left[ z~\sum_{\ell=0}^J (2\ell +1)~{e^{-\beta \sqrt{{\ell(\ell+1)\over R^2}}}\over \beta^3 R^2}\left(\beta^2 \ell(\ell+1)+ 2\left(1+\sqrt{\ell(\ell+1)\over R^2}\right)R^2\right)\right.\hspace{1cm}$$
$$\hspace{2cm} \left.\mp z^2~\sum_{\ell=0}^J (2\ell +1)~{e^{-2\beta \sqrt{{\ell(\ell+1)\over R^2}}}\over8 \beta^3 R^2}\left(4\beta^2 \ell(\ell+1)+ 2\left(1+\sqrt{\ell(\ell+1)\over R^2}\right)R^2\right)\right],\eqno{(6.11)}$$
in which $S$ is the area of Minkowski space.  Using the Euler-Maclaurin formula in (2.4) to perform the summation in above the total particle number becomes 
$$ N \approx {2\pi S\over h^2}\left[z\left(6R^2 (kT)^4 - 2J^2 (kT)^2 e^{-J(J+1)\over RkT}+{1\over 2}(kT)^2\right)\right.\hspace{4cm}$$
$$\left. \mp z^2 \left(6R^2 (kT/2)^4 - 2J^2 (kT/2)^2 e^{-J(J+1)\over R^2kT/2}+{1\over 2}(kT/2)^2 \right)\right],\eqno{(6.12)}$$
and the total energy becomes
$$ E \approx {2\pi S\over h^2}\left[z\left(24R^2 (kT)^5 - 10J^2 (kT)^3 e^{-J(J+1)\over RkT}+(kT)^3\right) \right.\hspace{4cm}$$
$$\mp \left.z^2\left(24R^2 (kT/2)^5 - 10J^2 (kT/2)^3 e^{-J(J+1)\over RkT/2}+(kT/2)^3\right) \right],\eqno{(6.13)}$$
Using (6.12) we can express the fugacity $z$ as a function of $N$.  After substituting this relation into (6.13) we can find the energy and the associated heat capacity at high temperature becomes
$$ C\approx ~k~\left(2+ {J^2\over 2R^2} {1\over (kT)^2}\right) n_0  \pm~k~ \left({1\over 16}+ {13\over192}{J^2\over R^2} {1\over (kT)^2}\right) n_0^2,\eqno{(6.14)}$$
in which ${n_0}$ is the particle number density.  Above result shows that the heat capacity of   fermion is a decreasing function of temperature, in high-temperature limit, contrast to that always being an increasing function on the commutative geometry [12].  Above property also shows in 1+3 Minkowski spacetime with extra fuzzy sphere, after numerical analysis. 

     According to the equipartition theorem [12] the particle on the 1+2 Minkowski spacetime will have two degrees of freedom.  As each degree of freedom of the relativistic particle will contribute the heat capacity $k~n_0$ in high-temperature limit.  Result in (6.14) tells us that the relativistic particle in 1+2 Minkowski spacetime with extra fuzzy sphere behaves as that on 1+2 Minkowski spacetime in high-temperature limit.  This shows explicity the ``mechanism of  thermal reduction of the fuzzy space".   

\section {Discussion}
In this paper we have studied the thermodynamics of ideal gas on the spacetime with extra fuzzy geometry.   We first evaluate the heat capacities of the non-relativistic ideal boson and fermion on the fuzzy two-sphere.  We see that they have different values, contrast to that on the commutative geometry [12].  We have calculated the ``statistical  interparticle potential"  [13] and see that  the noncommutativity of the fuzzy sphere has an inclination to enhance the statistical ``attraction (repulsion) interparticle potential" between boson (fermion).   This statistical  property may be used to explain why the ideal boson and fermion on the fuzzy two-sphere have different value of heat capacity.  We also see that, at high temperature the heat capacity approaches to zero as the all quantum levels on fuzzy two-sphere are occupied.

     We next evaluate the heat capacity of the non-relativistic ideal boson and fermion on the product of  1+D (with D=2,3) Minkowski spacetime by a fuzzy two-sphere and see that the heat capacity is a decreasing function of temperature in high-temperature limit.  We argue that at high temperature the quantum level of  extra fuzzy space is all occupied and the particle will behave as that on the a reduced space of 1+D Minkowski spacetime .  This ``mechanism of  thermal reduction of the fuzzy space" could lead the heat capacity of boson and fermion to be a decreasing function of temperature. 

     We finally investigate the relativistic system and evaluate the heat capacity of the free scalar and Dirac field on the product of  1 + 2 Minkowski spacetime by a fuzzy two-sphere.  We also find the similar properties in the relativistic system. 
\newpage
\begin{center} {\bf REFERENCES}\end{center}
\begin{enumerate}
\item H. S. Snyder, Phys. Rev. 71 (1947) 38; 72 (1947) 68.
\item Connes, A. Connes, ``Noncommutative Geometry", Academic Press, New York, 1994.;\\ Connes, ``A.Gravity coupled with matter and the foundation of non commutative geometry, Comm. in Math. Phys. 182 155-177 (1996), hep-th/9603053 .
\item   M. R. Douglas and N. A. Nekrasov, ``Noncommutative Field Theory ",  Rev.Mod.Phys.73 (2001) 977 [hep-th/0106048];\\
R. J. Szabo, Quantum Field Theory on Noncommutative Spaces, Phys. Rept. 378 (2003) 207 [hep-th/0109162]. 
\item  A. P. Balachandran, S. Kurkcuoglu and S. Vaidya, ``Lectures on Fuzzy and Fuzzy SUSY Physics ", [hep-th/0511114].
\item A. H. Chamseddine and Al. Connes,  ``Formula for Noncommutative Geometry Actions: Unification of Gravity and the Standard Model", Phys. Rev. Lett. 77 (1996) 4868; \\A. H. Chamseddine, G. Felder, J. Frohlich ``SO(10) Unification in Non-Commutative Geometry ", Phys.Rev. D50 (1994) 2893[hep-th/9304023 ]; \\A. Connes, "Noncommutative Geometry and the Standard Model with Neutrino mixing", JHEP 0611(2006)081 [arXiv:hep-th/0608226];\\ A. H. Chamseddine, A. Connes and M. Marcolli, "Gravity and the Standard Model with Neutrino Mixing", Adv. Theor. Math. 11 (2007) 991 [hep-th/0610241 ]; \\
J. Barrett, "The Lorentzian Version of the Noncommutative Geometry Model of Particle Physics", J. Math. Phys. 48(2007)012303 [hep-th/0608221]. 
\item  A. Connes, M. R. Douglas and A. Schwarz, ``Noncommutative Geometry and Matrix Theory: Compactification on Tori", JHEP 9802:003 (1998) [hep-th/9711162]. 
\item  N. Seiberg and E. Witten, ``String Theory and Noncommutative Geometry", JHEP 9909 (1999) 032 [hep-th/9908142]. 
\item  A. Y. Alekseev, A. Recknagel and V. Schomerus, ``Non-commutative world-volume geometries: Branes on su(2) and fuzzy sphere" JHEP 9909 (1999) 023 [hep-th/9908040]; ``Brane dynamics in background  and non-commutative geometry," JHEP 0005 (2000) 010 [hep-th/0003187];
 \\Pei-Ming Ho, ``Fuzzy Sphere from Matrix Model" JHEP 0012 (2000) 015 [hep-th/0010165];\\
K. Hashimoto and K. Krasnov, ``D-brane Solutions in Non-Commutative Gauge Theory on Fuzzy Sphere," Phys. Rev. D64(2001) 046007 [hep-th/0101145];\\
S. Iso, Y. Kimura, K. Tanaka, and K. Wakatsuki,"Noncommutative Gauge Theory on Fuzzy Sphere from Matrix Model," Nucl. Phys. 604(2001) 121 [hep-th/0101102].  
\item  R. C. Myers, ``Dielectric-branes," JHEP 9912 (1999) 022 [hep-th/9910053];\\
C. Bachas, M. Douglas and C. Schweigert, ``Flux stabilization of D-branes,"  JHEP
0005 (2000) 048 [hep-th/0003037]. 
\item   H. Grosse, C. Klimcik and P. Presnajder, ``Towards finite quantum field theory in noncommutative geometry," Int. J. Theor. Phys. 35 (1996) 231  [hep-th/9505175]; ``On Finite 4D Quantum Field Theory in Non-Commutative Geometry", Commun.Math.Phys. 180 (1996) 429 [hep-th/9602115];\\
J. Gomis, T. Mehen and M.B. Wise, ``Quantum Field Theories with Compact Noncommutative Extra Dimensions," JHEP 0008, 029 (2000) [hep-th/0006160]. 
\item  S. Vaidya, ``Perturbative dynamics on fuzzy $S^2$ and $RP^2$", Phys.Lett. B512 (2001) 403 [hep-th/0102212]; \\C. S. Chu, J. Madore and H. Steinacker, ``Scaling Limits of the fuzzy sphere at one loop", JHEP 0108 (2001) 038  [hep-th/0106205];\\
B. P. Dolan, D. O. Connor and P. Presnajder, ``Matix $\phi^4$ models on the fuzzy sphere and their continuum limits", JHEP 0203 (2002) 013 [hep-th/0109084];\\
D. P. Jatkar, G. Mandal, S. R. Wadia, K. P. Yogendran, ``Matrix dynamics of fuzzy spheres",  JHEP 0201 (2002) 039 [ hep-th/0110172].
\item  R. K. Pathria,``Statistical Mechanics", (Pergamon, London, 1972).
\item  G. E. Uhlenbeck and L. Gropper, Phys. Rev. 41 (1932) 79.
 \item Wung-Hong Huang, "Effective Potential on Fuzzy Sphere", JHEP 0207 (2002) 064 [hep-th/0203051]; ``Quantum Stabilization of Compact Space by Extra Fuzzy Space", Phys.Lett. B537 (2002) 311 [hep-th/0203176]; ``Casimir Effect on the Radius Stabilization of the Noncommutative Torus", Phys.Lett. B497 (2001) 317 [ hep-th/0010160]; ``Finite-Temperature Casimir Effect on the Radius Stabilization of Noncommutative Torus ", JHEP 0011 (2000) 041 [hep-th/0011037]; Wung-Hong Huang and Kuo-Wei Huang, "Thermodynamics on Noncommutative Geometry in Coherent State Formalism", Phys. Lett. B670 (2009) 416 [arXiv:hep-th/0808.0324].  
\item  M. Abramowitz and I. A. Stegun, ``Handbook of Mathematical Functions with Formulas, Graphs, and Mathematical Tables" (New York, 1972). 
\end{enumerate}
\end{document}